# Calculation Energy levels and charge radius for odd $^{41\text{-}49}$Ca Isotopes by using the analytical approach

Mohsen Mousavi[1]*, Mohammad Reza Shojaei[1]
[1]Physics Department, Shahrood University of Technology, P.O. Box 3619995161-316, Shahrood, Iran
Email: nuclear.physics2020@gmail.com

**Abstract:** In the present study, some static properties of odd isotopes of Ca were investigated in the non-relativistic shell model. We also suggested a novel suitable local potential model for the non-microscopic investigation of the mentioned nuclei. We modeled the odd $^{41\text{-}49}$Ca nuclei as doubly-magic isotopes, with further nucleons (valence) in the $1f_{7/2}$ and $2p_{3/2}$ levels. Then the modified Eckart potential as well as Hulthen potential were chosen for the interaction between core and nucleons. We also used the Parametric Nikiforov–Uvarov method to calculate the values of energy, the radius of charge and wave function. The obtained results showed a good agreement with the experimental data, so this model is applicable for the similar nuclei.

## 1 Introduction

Few-body systems are dealt with in different fields of physics, including a large range such as very small structures or the large-scale gravitational systems. Since a long time ago, the equations of Schrodinger in non-relativistic and K-G Dirac in relativistic have been known as the crucial ways for investigating the atoms, nuclei, molecules and their spectral behavior, as well. A lot of researchers have tried to discover the most precise solution for bound state of these non-relativistic and relativistic equations for different potentials which clarify the nature of bonding or the vibration of quantum systems. Though, there is not an exact and practical solution to the Schrodinger equation. So, switching to other techniques like second quantization, quantum-field theory and Green's function would be effective. The second quantization is necessary to elucidate the creation and destruction phenomena of the particles in a relativistic theory. In a nonrelativistic theory, the second quantization unexpectedly eases the debate about lots of similar interacting components. The quantum field theory methods provide the possibility of focusing on few interesting matrix elements. In most of noticed cases, the first few orders of perturbation theory are not able to present a sufficient explanation for interacting N-particle system. Because of this, finding a systematic way for an analytical solution of the Schrodinger equation is crucial [1, 2, and 3].
Some of the static characteristics of the nucleus, e.g. levels of energy and radius of charge, are considered helpful for clarifying the structure of the nucleus [4]. Figuring out the forces between the nucleons, the nuclei structure, and the nature of the nuclear interactions between them and also with other subatomic particles, are considered as the main aims of studying nuclear physics. The emphasis of nuclear structure research is on studying the characteristics of the nuclei like exited states energies, nuclear shapes, electromagnetic moments and transition rates between the exited states and the ground states, and the way in which the nuclei change to other nuclei. The studies on nuclear structure reveal some facts about the individual nuclei's structure, as well as some experimental information which can be applied by the theorists to elucidate the nature of the nuclear forces [5]. Due to the valuable experimental results obtained about the binding energy, distribution of density, single particle energy, radius and etc., calcium isotopes have been in the center of attention. Calculating these quantities to further study the microscopic and non-microscopic theories in the latter experiments will be beneficial [6, 7]. There are 25 isotopes for calcium, including $^{34}$Ca to $^{58}$Ca [8]. The best proof of the behavior of single-particle can be obtained by the magic (also known as closed-shell) nuclei, where the quantity of protons and/or neutrons of the nucleus fills the last shell prior to a major or minor shell gap [9].

In the present study, the non-relativistic shell model is used to study the energy levels and charge radius of the odd $^{41\text{-}49}$Ca isotopes. Due to the fact that these isotopes have some nucleons out of the core ($^{40}$Ca), equation of Schrodinger is applied to investigate them in non-relativistic shell model. The nucleon and the core are considered as structure-less particles in the non-microscopic few-body model. Then, a proper potential is proposed for the interaction between the extra nucleon and the core, and a mathematical method is suggested to calculate the energy of these nuclei as non-relativistic systems. It is also mentioned above that the levels of the energy of the system is a very crucial parameter to further study these nuclei. Thus, the focus of this article is calculating this important parameter. The suggested mathematical model is discussed thoroughly in the subsequent sections, also, it will be proved that the obtained results are in a good agreement with the previous experimental reports and theoretical data.

## 2 Proposed Non-Relativistic Analytical Method

We should first solve the time-independent Schrödinger equation for calculating the energy of a multi-body system from a non-relativistic point of view in the presence of that kind of local potential which is just a function of the inter-particle distance ($x$).

We define a set of Jacobi coordinates for which $\zeta_N = r_{ij}$, Where $r_{ij}$ stands for the set of relative coordinates of the particle $r_{ij} = r_j - r_i$ . The center of mass R can be eliminated by using the Jacobi coordinates:

$$\zeta_i = \sqrt{\frac{i}{i+1}}\left(r_{i+1} - \frac{1}{i}\sum_{j=1}^{i} r_j\right), \qquad R = \frac{1}{N}\sum_{i}^{N} r_i\,,\ \ i = 1,\dots,N-1 \tag{1}$$

Where

$$x^2 = \sum_{i=1}^{N-1}\left(\zeta_i^2\right), \qquad t_i = \arctan\left(\frac{\left(\sum_{j=1}^{i}\left(\zeta_j^2\right)\right)^{1/2}}{\zeta_{i+1}}\right); \tag{2}$$

For example, a three-particle system after eliminating the center-of-mass motion becomes a six-dimensional one (D=9-3=6). From Eqs. (1) and (2), the internal three-body system can be well described by two Jacobi relative coordinates $\zeta_1$ and $\zeta_2$ and R=R$_3$ defined as

$$\zeta_1 = \frac{r_1 - r_2}{\sqrt{2}}, \qquad \zeta_2 = \frac{r_1 + r_2 - 2r_3}{\sqrt{6}}, \qquad R_3 = \frac{r_1 + r_2 + r_3}{3} \tag{3}$$

Instead of $\zeta_1$ and $\zeta_2$, one can introduce the hyper-spherical coordinates, with the hyper-radius $x$ and the hyper-angle $t$ described by:

$$x = \sqrt{\zeta_1^2 + \zeta_2^2}, \qquad t = \arctan\left(\frac{\zeta_1}{\zeta_2}\right) \tag{4}$$

In the provided method for solving Schrödinger equation for a system of N-identical particles with a hyper-spherical formalism, the part of Schrödinger equation related to the hyper-radius $x$ can be demonstrated as follows [10, 11].

$$\left\{\frac{d^2}{dx^2} + \frac{D-1}{x}\frac{d}{dx} - \frac{\ell(\ell + D - 2)}{x^2} - \frac{2m}{\hbar^2}\left[V(x) - E_{n\ell}\right]\right\} R_{n\ell}(x) = 0 \tag{5}$$

In which $R_{n\ell}(x)$, $E_{n\ell}$ *and* $\ell$ represent the hyper-radial part, the Eigen-values of energy, and the orbital angular momentum, respectively. The *N*-body problem in the center-of-mass frame is mathematically (*3N-3*)-dimensional. According to the hyper-spherical method, we can claim that if there is a point in the (*D=3N-3*)-dimensional configuration space, it can be presented as lying on a (*D-1*)-dimensional hyper-sphere which its radius is $x$ [12].

In nuclear physics, Eckart potential and Hulthen potential can be mentioned among the most applicable potentials for investigating bound states and scattering parameters. The hyper-spherical harmonic formalism can introduce and treat the many-body forces very easily. The definition of modified Eckart potential plus Hulthen potential is shown below [13, 14],

$$\mathrm{V(x)} = \mathrm{v}_0 \operatorname{cosech}^2(\alpha \mathrm{x}) + \mathrm{v}_1 \frac{\mathrm{e}^{-2\alpha \mathrm{x}}}{(1-\mathrm{e}^{-2\alpha \mathrm{x}})} = 4\mathrm{v}_0 \frac{\mathrm{e}^{-2\alpha \mathrm{x}}}{(1-\mathrm{e}^{-2\alpha \mathrm{x}})^2} + \mathrm{v}_1 \frac{\mathrm{e}^{-2\alpha \mathrm{x}}}{(1-\mathrm{e}^{-2\alpha \mathrm{x}})} \tag{6}$$

Where $v_0$ and $v_1$ stand for the nuclear repulsion and attraction strengths, respectively, and α is represents a parameter which is related to the potential range. In the suggested potential, the first and second terms are of Eckart and Hulthen types, respectively.

Eq. (7) can be obtained from the Eq. (5) by considering that radial wave function is as $U(x) = x^{(D-1/2)} R_{n\ell}(x), \quad \lambda = \ell + \frac{(D-3)}{2}$ In which the hyper-central potential is derived from the two-body potential.

$$\frac{d^2 U_{n\ell}(x)}{dx^2} + \frac{2m}{\hbar^2}\left\{E_{n\ell} - 4v_0 \frac{e^{-2\alpha x}}{(1-e^{-2\alpha x})^2} - v_1 \frac{e^{-2\alpha x}}{(1-e^{-2\alpha x})} - \frac{\hbar^2}{2m}\frac{\lambda(\lambda+1)}{x^2}\right\} U_{n\ell}(x) = 0 \tag{7}$$

Eq. (7) can be exactly solved only if $\lambda=0,-1$. For this reason, an approximation should be used to deal with the centrifugal-like terms. For the purpose of gaining the analytical solutions of Eq. (7), the improved Greene and Aldrich [15] approximation is used and the spin-orbit coupling term is replaced with the expression which is valid for $\alpha x<<1$ [16]. The main property of the mentioned solutions is resulted by the replacement of the centrifugal term by an approximation one so that a solvable equation, normally hyper-geometric, is obtained [17].

$$\frac{\lambda(\lambda+1)}{x^2} \approx \frac{\lambda(\lambda+1)4\alpha^2 e^{-2\alpha x}}{\left(1-e^{-2\alpha x}\right)^2} \tag{8}$$

Fig. 1 shows the behavior of the improved approximation. A good agreement can be observed for small values of *α*.

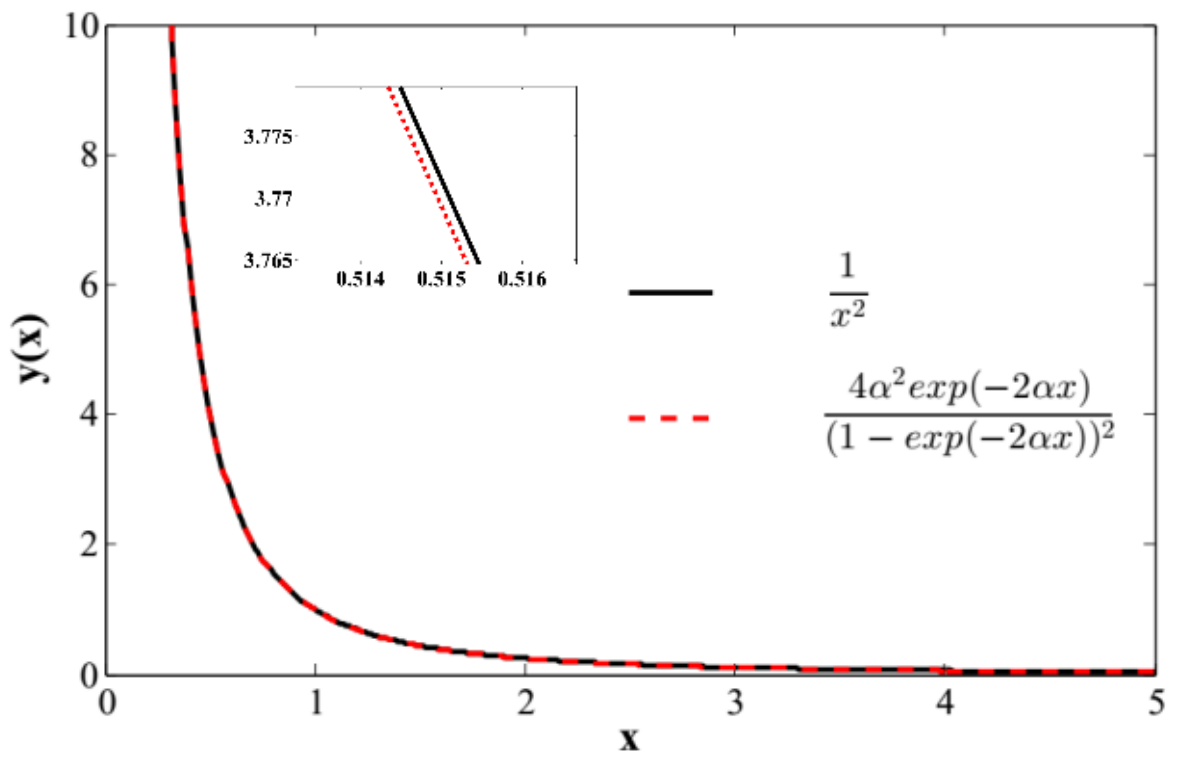

**Fig. 1.** The behavior approximation for $\alpha = 0.08\ \text{fm}^{-1}$.

Eq. (7) can be summarized as follows, after applying the approximations of Eq. (8) for a centrifugal term and hyper-central potential, as well as inserting a new variable $s=\exp(-2\alpha x)$,

$$U''_{n,\ell}(s)+\frac{(1-s)}{s(1-s)}U'_{n,\ell}(s)+\frac{1}{s^2(1-s)^2}\left[-\chi_2 s^2+\chi_1 s-\chi_0\right]U_{n,\ell}(s)=0 \tag{9}$$

Where $\chi_2$, $\chi_1$ and $\chi_0$ parameters are considered as mentioned below:

$$\chi_2=-\frac{m}{2\alpha^2\hbar^2}\left[E_{n\ell}+v_1\right]$$

$$\chi_1=-\frac{m}{\alpha^2\hbar^2}\left[E_{n\ell}+2v_0+\frac{v_1}{2}\right]-\lambda(\lambda+1) \tag{10}$$

$$\chi_0=-\frac{m}{2\alpha^2\hbar^2}E_{n\ell}$$

As it is perceivable, the Schrödinger equation with the considered potential for the interaction of the nucleon and the close shell has changed to a second-order differential equation which is in the agreement with the general form of the equation in Parametric Nikiforov–Uvarov (PNU) method. This method can solve Eq. (9). PNU method can be used to acquire the equation of energy (referring to the references [18, 19]) as:

$$(2n+1)\left(\sqrt{\chi_2-\chi_1+\chi_0+\frac{1}{4}}+\sqrt{\chi_0}+\frac{1}{4}(2n+1)\right)+2\sqrt{\chi_0(\chi_2-\chi_1+\chi_0+\frac{1}{4})}+2\chi_0-\chi_1+\frac{1}{4}=0 \tag{11}$$

In the last step, the energy equation can be gained as follows, by considering the notations of Eqs. (10) and (11):

$$E_{n\ell}=-\frac{2\alpha^2\hbar^2}{m}\left\{\frac{\frac{[2n+2\gamma+1]^2}{4}+\frac{mv_1}{2\alpha^2\hbar^2}}{[2n+2\gamma+1]}\right\}^2;\gamma=\left[\frac{2m}{\alpha^2\hbar^2}v_0+\lambda(\lambda+1)+\frac{1}{4}\right] \tag{12}$$

And we can present the hyper-radial wave function, with referring to PNU method in references [20, 21], as:

$$R_{n,\ell}(x)=N'x^{-\left(\frac{D-1}{2}\right)}\left(e^{-2\alpha x}\right)^{\left(\sqrt{\chi_0}\right)}\left(1-e^{-2\alpha x}\right)^{\left(\frac{1}{2}+\sqrt{\chi_2-\chi_1+\chi_0+\frac{1}{4}}\right)}P_n^{\left(2\sqrt{\chi_0},2\sqrt{\chi_2-\chi_1+\chi_0+\frac{1}{4}}\right)}\left(1-2e^{-2\alpha x}\right) \quad (13)$$

Where $N'$ is the constant of normalization and the $P_n^{(\alpha,\beta)}(x)$ functions represent the Jacobi polynomials.

In figure 2, binding energy of the ground state was studied by making the refrence to Eq. (12) for two, three and four-body systems versus the varying amounts of $v_1$ and $v_0$ for a fixed amount of $\alpha=0.08fm^{-1}$. It is obvious that by increasing $v_0$ and $v_1$, the binding energy of the ground state of the system increases and decreases, respectively.

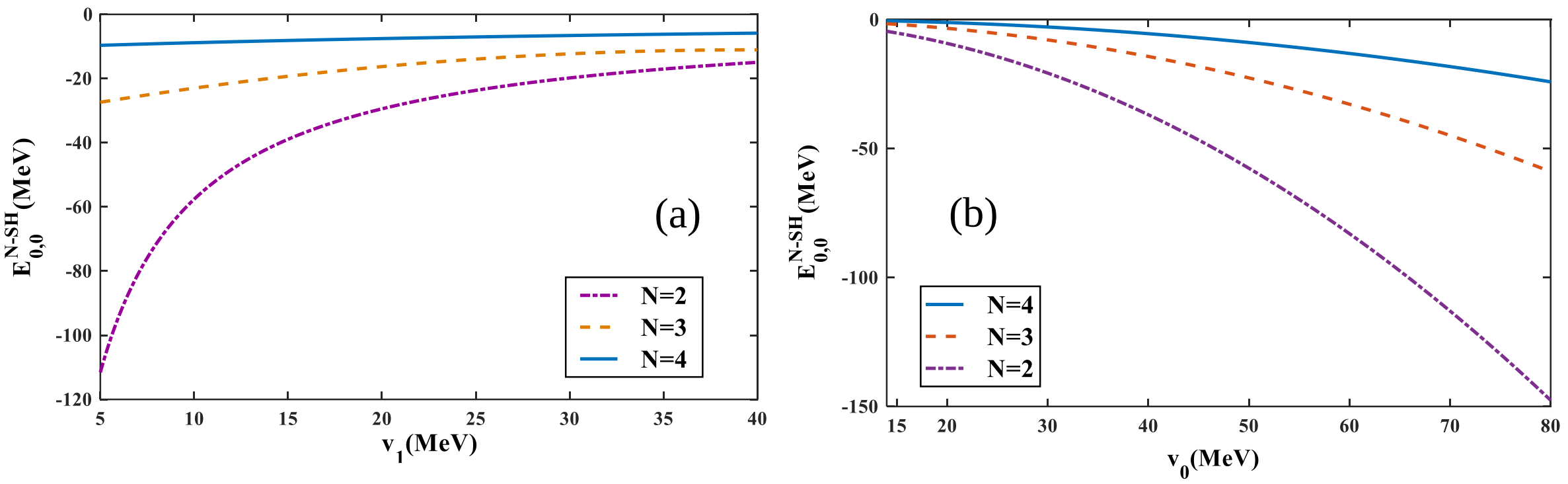

**Fig. 2** Comparison between of the Schrodinger ground state energy for two-, three- and four-body systems versus the different value of (**a**) $v_1$ and (**b**) $v_0$ for a fixed value of $\alpha$=0.08 fm$^{-1}$.

A very important challenge in nuclear structure is to realize the shell structure development in all details, from the valley of stability to neutron-rich extremes [22, 23]. If the proton shell is closed, an ideal region will be provided by calcium isotopes for investigating the formation and development of the shell in medium-mass nuclei from nuclear forces. After acquiring the substantial Eq. 12, the energy levels of $^{41-49}$Ca isotopes can be calculated if we assign proper values for the coefficients of the potential equation. The $^{41-49}$Ca isotopes with nucleons are considered to be on top of the $^{40}$Ca isotope core. For instance, the $^{43}$Ca, $^{45}$Ca and $^{47}$Ca nuclei may be modeled as a doubly magic $^{40}$Ca (N=Z=20), with additional (valence) 3, 5 and 7 nucleons in the $lf_{7/2}$ level, respectively. The spin and parity of the ground state of these isotopes are $J^{\pi}=7/2^-$, which is related to the spin and parity of the level that the valence nucleon occupies. Calcium 41 and 49 can be examined in different method. This method includes a doubly-magic $^{40}$Ca (N=Z=20) and $^{48}$Ca (N=28, Z=20) with one nucleon in the $lf_{7/2}$ and $2p_{3/2}$ levels, respectively. The ground state and excited states energies of $^{41,49}$Ca isotopes are obtained by using Eq. (12) and Eq. (13) (with a few change in Eq. (12) by D=3, $\lambda\rightarrow l$). So, the energies of the ground state and the excited states of $^{41-49}$Ca isotopes can be acquired through Eq. (12). The spin of the nucleus at the ground state is $J^{\pi}=3/2^-,7/2^-$, so the impact of the spin-orbit coupling, using equation (14), on the levels of the energy is computed as a first-order disturbing factor. The values of potential parameters are shown in table 1.

$$E^{(1)}_{n,\ell} = \langle n | V_{L.S}(x)\vec{L}.\vec{S} | n\rangle = \int R^{(0)*}_{n,\ell}(x) \frac{\hbar^2}{2m_0^2c^2}\frac{1}{x}\frac{dV(x)}{dx}\vec{L}.\vec{S}\, R^{(0)}_{n,\ell}(x)\, x^2 dx \qquad (14)$$

In this equation, $R^{(0)}_{n,\ell}(x)$ represents the undisturbed wave function which is obtained from equation (13), $m_0$ is the mass of the nucleon and c is speed of the light [24]. We compared these results with the experimental data, as summarized in figure 3 [25].

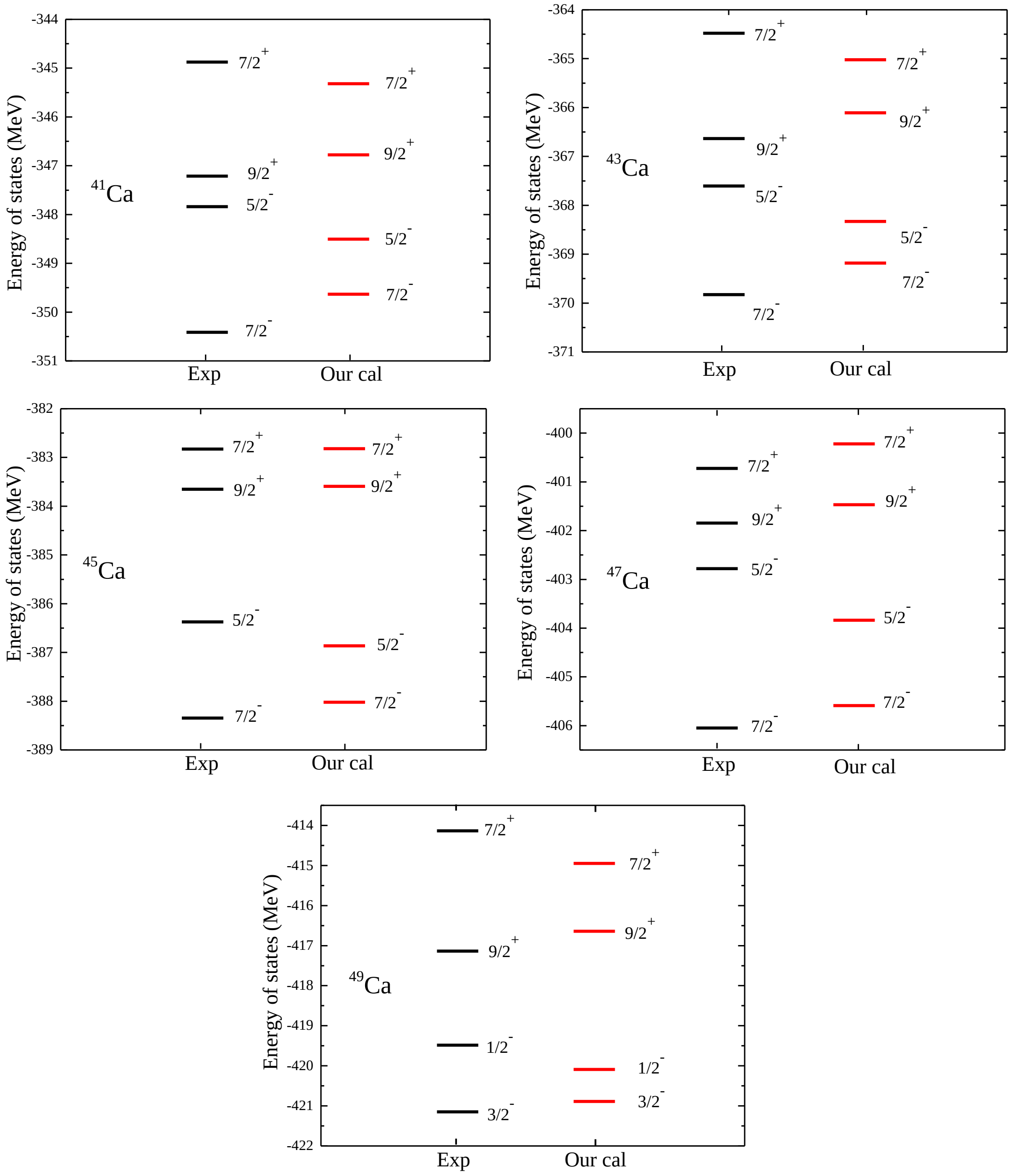

**Fig. 3.** Comparison between calculated levels and experimental data for $^{41\text{-}49}$Ca isotope.

The calculated energy levels are close to the experimental data. Consequently, the suggested model can also be used for investigating other similar isotopes.

The radial wave function is obtained from equation (13), so we can easily calculate the charge radius of studied isotopes by calculating $<x^2>^{1/2}$. The obtained results are shown in Table 3.

$$\langle x^2\rangle^{\frac{1}{2}} = \left( \frac{\int R^*_{n,\ell}(x) x^2 R_{n,\ell}(x) d^3x}{\int R^*_{n,\ell}(x) R_{n,\ell}(x) d^3x} \right)^{\frac{1}{2}} \qquad (15)$$

**Table 1:** the charge radius (fm) for isotopes for ground state energy.

| Isotope | Parameters of potential | | | $< x^2 >^{\frac{1}{2}}_{our\ work}\ (fm)$ | $< x^2 >^{\frac{1}{2}}_{EXP}\ (fm)$ [26] |
|---|---|---|---|---|---|
| | **α(fm$^{-1}$)** | **V$_0$ (MeV)** | **V$_1$ (MeV)** | | |
| **$^{41}$Ca** | 0.0244 | 2.3266 | -142.2208 | 3.7614 | 3.4780 |
| **$^{43}$Ca** | 0.0246 | 4.0268 | -198.1047 | 4.006 | 3.4954 |
| **$^{45}$Ca** | 0.0247 | 5.4867 | -246.1046 | 3.9730 | 3.4944 |
| **$^{47}$Ca** | 0.0249 | 6.2917 | -282.8943 | 3.8103 | 3.4783 |
| **$^{49}$Ca** | 0.0254 | 3.3361 | -192.2486 | 3.7909 | ----- |

As comparison between these results and experimental data implies [26], the suggested model can estimate the charge radius of these isotopes in their ground states.

## 3 Conclusions

In this study, the problem of non- relativistic few-body bound system was investigated by presenting the analytical solution of D-dimensional Schrödinger equation. To obtain the energy Eigen-values and wave functions for a few-body bound system the Parametric Nikiforov-Uvarov method is applied. Fig. 2 shows the effect of the potential parameters of the non- relativistic systems with two, three and four nucleons on the binding energies. The offered approach here can be applicable in surveying the non-relativistic corrections pertaining to the observables characterizing the characteristic of few-body nuclear systems, within a simple treatment.

As figure 3 and table 1 present, the levels of energy and radius of the charge, computed using the suggested potential in this paper, show a good compatibility with the experimental data for the intended nuclei that are referred to as preliminary examples. The principal aim of the present survey is obtaining a united method for applying the analytical model for all of resembling nuclei. So, by utilizing a suitable non-microscopic method, these nuclei can be assumed as few-body structures. Eventually, it is deduced that the suggested potential and mathematical model may be suitable for other equivalent nuclei in the non-microscopic approach for the $^{41\text{-}49}$Ca isotopes. More theoretical studies should be performed for successful clarification of the application of this method for different properties of similar nuclei.

## Reference

[1] R. P. Feynman, Phys. Rev. **76**, 749 (1949).

[2] F. J. Dyson, Phys. Rev. **75**, 486 (1949).

[3] M. M. Giannini, E. Santopinto, A.Vassallo, Prog. Part. Nucl. Phys. **50**, 263 (2003).

[4] Z Shuangquan, M Jie and Z Shangui, *Sci. China Phys. Mech. Astron.* **46**, 632 (2003).
[5] M.G. Mayer, J.H.D. Jensen, Elementary Theory of Nuclear Shell Structure, Wiley, New York, 1955.
[6] E.H. Auerbach, S. Kahana, J. Weneser, Phys. Rev. Lett. 23 (1969) 1253.
[7] G. Audi, A.H. Wapstra, C. Thibault, J. Blachot, O. Bersillon, Nucl. Phys. A729 (2003) 3–128.
[8] http://www.nndc.bnl.gov/.
[9] B.L. Cohen, Concepts of Nuclear physics, McGraw-Hill, New York, 1971.
[10] A. A. Rajabi, Few-Body Systems, **37** 197—213 (2005).
[11] U. A. Deta, Suparmi and Cari, Adv. Studies Theor. Phys. 7 647 (2013).
[12] J. Avery, Hyperspherical Harmonics: Applications in Quantum Theory (Dordrecht:Kluwer, 1989).
[13] M.R. Shojaei and M. Mousavi, Advances in High Energy Physics, vol. 2016, Article ID 8314784, 12 pages, (2016).
[14] B.J. Falaye. Cent. Euro. J. Phys 10 960 (2012).
[15] R. L. Greene, C. Aldrich, Phys. Rev. A **14** 2363 (1976).
[16] M. Mousavi and M. R. Shojaei, *Commun. Theor. Phys.* **66** 483 (2016).
[17] F. J. S. Ferreira, F. V. Prudente, Physics Letters A, **377**, 3027-3032 (2013).
[18] M. Mousavi, M. R. Shojaei, Chin. J. Phys. **54**, 750–755 (2016).
[19] M. Mousavi and M. R. Shojaei, *Pramana - J Phys* **88**: 21 (2017).
[20] M. R. Shojaei and N. Roshanbakht, Chin. J. Phys. **53**, 120301 (2015).
[21] H. Yanar and A. Havare, Adv. High Energy Phys. vol. 2015, Article ID 915796, 17 pages, (2015). doi:10.1155/2015/915796
[22] T. Baumann, A. Spyrou, and M. Thoennessen, Rep. Prog. Phys. **75**, 036301 (2012).
[23] O. Sorlin and M.-G. Porquet, Prog. Part. Nucl. Phys. **61**, 602 (2008).
[24] N. Roshanbakht and M. R. Shojaei, Advances in High Energy Physics, vol. 2017, 2017.
[25] G. Audi, A. H. Wapstra and C. Thibault, *Nucl. Phys. A*, **729** 3 (2003).
[26]. K. P. Angeli and Marinova . Atomic Data and Nuclear Data Tables, 99 ,69 (2013).